\documentclass{aa}
\usepackage{array}
\usepackage{graphicx}
\usepackage{lscape}
\usepackage{natbib}
\usepackage{longtable}
\usepackage{color}
\usepackage{mathrsfs} 
\usepackage{txfonts,textcomp}
\usepackage{tikz}
\usepackage{amssymb}
\usepackage{multirow}
\usepackage{pdflscape}
\usepackage{txfonts}

\begin{document}

\title{Isochrone-cloud fitting of the extended Main-Sequence Turn-Off of 
young clusters\thanks{Isochrone-cloud files are available to the astronomical
  community from CDS...}}

\author{C. Johnston\inst{1}
  \and C. Aerts\inst{1,2,3}
  \and M.~G. Pedersen\inst{1}
  \and N. Bastian\inst{4}}

\institute{Instituut voor Sterrenkunde, KU Leuven, Celestijnenlaan 200D, 3001
  Leuven, Belgium, \email{colecampbell.johnston@kuleuven.be}
  \and Department of Astrophysics, IMAPP, Radboud University Nijmegen, 
P. O. Box 9010, 6500 GL Nijmegen, the Netherlands
  \and Max Planck Institute for Astronomy, Koenigstuhl 17, 69117 Heidelberg,
  Germany
\and
Astrophysics Research Institute, Liverpool John Moores University, 
146 Brownlow Hill, Liverpool L3 5RF, UK
}

\date{Received Date Month Year / Accepted Date Month Year}

\abstract{Extended main-sequence turn-offs (eMSTO) are a commonly observed
  property of young clusters. A global theoretical interpretation for the eMSTOs
  is still lacking, but stellar rotation is considered a necessary ingredient to
  explain the eMSTO.}  {We aim to assess the importance of core-boundary and
  envelope mixing in stellar interiors for the interpretation of eMSTOs in terms
  of one coeval population.}  {We construct isochrone-clouds based on interior
  mixing profiles of stars with a convective core calibrated from asteroseismology 
  of isolated galactic field stars. We fit these isochrone-clouds to the measured
  eMSTO to estimate the age and core mass of the stars in the two young clusters 
  NGC1850 and NGC884, assuming one coeval population and fixing the metallicity 
  to the one measured from spectroscopy.  We assess the correlations between the
  interior mixing properties of the cluster members and their rotational and
  pulsational properties.}  {We find that stellar models based on
  asteroseismically-calibrated interior mixing profiles lead to enhanced core
  masses of eMSTO stars and can explain a good fraction of the observed 
  eMSTOs of the two considered clusters in terms of one coeval population 
  of stars, with similar ages to those in the literature, given the large 
  uncertainties.  The rotational and pulsational properties of the stars in
  NGC\,884 are not sufficiently well known to perform asteroseismic modelling,
  as it is achieved for field stars from space photometry.  The stars in
  NGC\,884 for which we have $v\sin i$ and a few pulsation frequencies show no
  correlation between these properties and the core masses of the stars that set
  the cluster age.}{Future cluster space asteroseismology may allow to interpret 
  the values of the core masses in terms of the physical processes that cause 
  them, based on the modelling of the interior mixing profiles for the individual 
  member stars with suitable identified modes.}

\keywords{Asteroseismology -- 
Stars: interiors --
Stars: oscillations (including pulsations) --
Stars: rotation --
Galaxies: clusters: general --
Galaxies: clusters: individual: NGC\,884, NGC\,1850
}

\titlerunning{Isochrone-cloud fitting of clusters}
\authorrunning{C. Johnston et al.}
\maketitle 
%%%%%%%%%%%%%%%%%%%%%%%%%%%%%%% Main text %%%%%%%%%%%%%%%%%

\section{Introduction}
\label{section:intro}

The theoretical understanding of the physical processes in stellar interiors and
their implementation in modern Stellar Structure and Evolution (SSE) codes are
fundamental cornerstones of astrophysics.  To this end, stellar astrophysics has
benefited tremendously from the exponential development of {\it
  asteroseismology}, i.e., the investigation of stellar interiors through the
characterisation and modelling of stellar oscillation modes. Major progress was
achieved the past decade thanks to space-based photometric missions
such as CoRoT, {\it Kepler}, BRITE, K2, and TESS.

One of the great benefits from these long time-base, high duty-cycle
high-precision observations has been the probing of the deep stellar interior of
stars born with a convective core. This can be achieved by asteroseismology of
long-period (order of days) gravity-mode oscillations (g-modes hereafter), which
was impossible from ground-based data, for stars in the core-hydrogen burning
phase. The g-modes in intermediate-mass stars with
$M_\star\in [1.3,8.5]$\,M$_\odot$ propagate between the stellar surface and the
convective core.  Such resonant g-modes are triggered by a flux-blocking
mechanism at the bottom of the convective envelope for the F-type
$\gamma\,$Doradus stars \citep{Dupret2005,Bouabid2013} and by a heat ($\kappa$)
mechanism in the outer radiative envelope acting upon the partial ionisation
layers or iron-like elements for slowly pulsating B (SPB) stars
\citep{Miglio2007,Szewczuk2017}.  Gravity-mode asteroseismology requires
uninterrupted monitoring of tiny brightness variations at $\mu$mag level during
several months and was first achieved from 5-months long CoRoT light curves
\citep{Degroote2010,Papics2012}.

The amount of mass in the convective core drives the evolution of intermediate
and high-mass stars and determines their age at the end of a given nuclear 
core-burning phase. It is therefore of utmost importance to estimate the 
amount of mixing that occurs in the transition layers between the convective 
core and the bottom of the radiative envelope. Gravity-modes allow to probe 
this particularly important region. Exploitation of detected g-mode oscillations 
revealed that stars rotate nearly rigidly throughout the main-sequence
\citep[MS,][]{Kurtz2014,Saio2015,Murphy2016,Aerts2017,VanReeth2018,LiG2019}.
This points to an efficient angular momentum (AM) transport mechanism that is
currently not included in standard SSE models. Such efficient AM transport and
near-rigid rotation not only occur during the MS but also during core-helium
burning, i.e.\ in phases when stars have a convective core \citep[e.g.,][for a
review]{Aerts2019}.

Both heat-driven resonant g-modes excited by the $\kappa$ mechanism and damped
travelling internal gravity waves (IGW) triggered stochastically at the
interface of the convective core and the radiative envelope have been suggested
as missing ingredients to remedy this discrepancy in the theory of AM transport
\citep{Rogers2013,Aerts2015,Rogers2015,Townsend2018}.  IGWs excited by the
convective core do not depend on metallicity but their propagation and
dissipation properties are poorly understood.  Recent CoRoT, K2, and TESS data
have revealed detection of IGWs in almost all OB stars, both in the Milky Way
and in the Large Magellanic Cloud \citep{Bowman2019a,Bowman2019b}. Hence, 
demonstrating that low metallicity stars, which do not undergo resonant 
heat-driven g-modes, can undergo efficient angular momentum transport via IGWs.

High-precision ageing, via asteroseismic estimation of the convective core mass,
only became possible from the 4-year long light curves assembled with the {\it
  Kepler\/} mission, which allowed to identify the detected g-mode frequencies
from their period spacing patterns.  The stellar age can be estimated from such
data by assessing the core mass of the star. This was done from measuring the
level of near-core boundary mixing (CBM), e.g., in the form of convective 
core overshooting, in single and binary B-type stars
\citep[e.g.,][]{Moravveji2015,Moravveji2016,Kallinger2017,Szewczuk2018,
  Johnston2019a} and in single F-type stars
\citep[e.g.][]{VanReeth2016,Mombarg2019}. The asteroseismically calibrated
levels of core overshooting led to values between typically 0.1 and 0.5 times
the local pressure scale height, resulting in increased core masses compared to
standard SSE models without extra mixing at the near-core boundary. With the
exception of one magnetic g-mode pulsator \citep{Buysschaert2018},
higher-than-standard core masses were found for almost all studied g-mode
pulsators of intermediate mass (some 50 stars), covering the entire
core-hydrogen burning phase, irrespective of their level of rotation.

The assumption that the core masses of intermediate-mass stars are solely due to
overshooting is a simplification. Indeed, in reality convective penetration by
plumes, overshooting, rotation, and oscillatory motions due to resonant modes
and IGWs all interplay and imply the occurrence of an overall net near-CBM 
profile. This overall CBM may further be affected (or not) by core magnetism 
\citep{Fuller2019} or by tides in the case of a close binary \citep{Song2013}. 
Moreover, chemical mixing also occurs in the radiatively stratified envelope. 
Such envelope mixing (hereafter denoted as REM) may be of either microscopic
or macroscopic nature. Inherently, these two forms of mixing produce opposite
effects in a given layer of the radiative envelope. Macroscopic mixing as 
triggered by rotation, (semi)convection, convective envelope boundary mixing 
\citep[via entrainment, penetration, or overshooting, see][]{Viallet2015}, waves, 
magnetism, tides, etc, occurs on relatively short timescales, and homogeneously 
mixes the layer. Alternatively, mixing of microscopic origin due to atomic diffusion
occurs on relatively long timescales. Here, we consider atomic diffusion as the
combined effect of gravitational settling, thermal diffusion, concentration 
diffusion, and radiative levitation \citep{Thoul1994,Michaud2015,Dotter2017}.
These processes can alter chemical gradients and create relative over/under-
abundances of given elements within the stably stratified radiative region.
The timescales involved in these processes determine which effect is dominant. 
If any large scale macroscopic mixing mechanism is active on a significantly
shorter timescale than atomic diffusion, the layer will be homogeneously 
mixed, effectively erasing the fingerprint of atomic diffusion. We assume
this to be the case in our models, such that the REM is only of macroscopic
origin. Unfortunately, the overall profile of neither the CBM nor the REM 
is well understood, as it cannot be derived from first principles 
\citep[e.g.,][for a review]{Salaris2017}. Recently, however, asteroseismic 
modelling of intermediate-mass stars has been able to probe the level of this 
CMB+REM mixing profile from g-mode period-spacing patterns and their deviations 
from uniformity \citep[e.g.,][]{Moravveji2015,Moravveji2016,SchmidAerts2016,Mombarg2019}.
Indeed, g-mode asteroseismology has been proven to be sensitive to the
morphology of the CBM+REM mixing profile within stars \citep{Pedersen2018}, as
well as to the temperature gradient within the CBM region adjacent to the
convective core \citep{Michielsen2019}.

Aside from g-mode asteroseismology, numerous studies which performed isochrone
modelling of binary stars have also indicated the need for enhanced core masses
\citep[e.g.,][among
others]{Ribas2000,Claret2007,Tkachenko2014,HiglWeiss2017,claret2019,Johnston2019b}.
However, binary modelling delivers less strong probes than g-modes and
uncertainties remain in the calibration of CBM using only binary constraints
\citep{ConstantinoBaraffe2018,Johnston2019a}. \citet{Johnston2019a} combined
binary  and g-mode asteroseismic modelling to show that derived core
mass estimates are in that case more constrained than CBM estimates from
binarity alone for a few double-lined g-mode pulsating binaries observed by {\it
  Kepler}.  \citet{Johnston2019b} made the same assertion for eclipsing
binaries, stressing that the core mass should be estimated and calibrated,
rather than the efficiency of a single mixing mechanism.

The asteroseismic modelling of individual single and binary field stars
discussed above has shown that mixing with a whole variety of levels and
profiles is active in intermediate-mass stars.  The major consequence of these
asteroseismically derived CBM and REM levels for stellar evolution is an
increase in the core mass of the star. As a result, asteroseismology has
provided a range of stellar core mass fractions along the MS, because
CBM brings fresh fuel into the core \citep[for a summary of CBM and REM values,
see the analyses by][for B- and F-type stars, respectively, hereafter jointly
referred to as MfA:
``Mixing-from-Asteroseismology'']{Briquet2007,Moravveji2015,Moravveji2016,SchmidAerts2016,Buysschaert2018,HendriksAerts2019,Mombarg2019,Johnston2019a}.
One of the most important implications of stars having a plethora of mixing
profiles and enhanced core masses is that the terminal-age MS (TAMS) is not a
single line in the Hertzsprung-Russel Diagram (HRD), but rather an entire region.
This has so far largely been ignored in cluster studies.

Extended main-sequence turn-offs (eMSTO) are ubiquitous in colour-magnitude
diagrams (CMDs) of young clusters in the Milky Way and in the Large and Small
Magellanic Clouds \citep[LMC and SMC,
e.g.,][]{Mackey2008,Milone2018,Marino2018a,Cordoni2018}.  The eMSTO is
characterised by an apparently broad TAMS region in the CMD, covering a given
width in colour, depending on the particular cluster.  Since its discovery, the
mechanism behind the eMSTO has been a matter of intense debate. Several
mechanisms have been proposed to explain the eMSTO phenomenon, such as age
spreads, binaries and binary interaction products, 
near-critical rotation, and convective penetration, with no single
mechanism being able to explain the full breadth of observed morphologies
\citep[e.g.][and references therein]{Li2019,Gossage2019}.   

The observed eMSTOs have been associated with fast surface rotation as revealed
by spectroscopy \citep[e.g.,][]{Dupree2017,Kamann2018,Marino2018b,Bastian2018}.
While some clusters with high-mass stars at the eMSTO show a correlation between
$H\alpha$ emission and location in the CMD (Bodensteiner et al., {\it submitted}),
others such as the double cluster $h$ and $\chi$ Persei reveal no relation
between colour and rotational velocities \citep{Li2019}. Furthermore, in
independent studies, both \citet{Li2019} and \citet{Beasor2019} demonstrate that
models consisting of a single population of stars with the same age and
metallicity but with a variety of rotation rates cannot explain the full
morphology of this double cluster and suggest that other mechanisms such as
binary interaction could be important.  While in general stellar evolution
models including rotational effects and magnetic braking offer improved
isochrone fits to eMSTOs in terms of coeval populations compared to non-rotating
non-magnetic models \citep{Georgy2019}, they fail to explain the full morphology
of the eMSTO in clusters of young to intermediate ages
\citep[e.g.,][]{Milone2018,Goudfrooij2018,Li2019}.  Whether this is caused by
limitations in the input physics used in the models, or the fitting methodology
is yet to be determined. \citet{Goudfrooij2018} pointed out that the MS
of young clusters are narrow for stellar masses below
$\sim\! 1.4\,$M$_{\odot}$ and broaden for all stars above this ``cut off''
mass. This minimal eMSTO mass coincides with the mass for which both a well
developed convective core and non-radial g-modes occur in MS stars
\citep[e.g.,][]{VanReeth2015,Papics2017}. As such, we investigate if the
asteroseismically calibrated mixing mechanisms discussed above, and particularly
their consequence in terms of increased core mass, play a role in the formation
of the eMSTO of young open clusters.

\citet{Yang2017} already considered the case of allowing for different levels of
convective penetration as the mechanism for CBM for non-rotating stellar models
of a coeval population and showed that this can explain the eMSTO to some
extent.  We generalise this approach to investigate to what level the mixing
profiles calibrated by asteroseismology of stars with a convective core can
explain the eMSTOs of young clusters. We do this by constructing so-called
isochrone-clouds first introduced by \citet{Johnston2019a}. These are areas in
the HRD covered by a coeval population of stars
having various masses and fractional core masses due to different interior
mixing profiles. These areas naturally form an extended TAMS region.  Unlike
\citet{Yang2017}, who only considered penetrative convection with un-calibrated
large values of the penetration distance up to 0.7 times the local pressure
scale height, we consider the general case of both CBM and REM and limit the 
overall amount of CBM and REM to measured levels calibrated by asteroseismology 
of intermediate- and high-mass field stars as achieved by the MfA studies. We fit
isochrone-clouds to CMDs of two prototypical clusters with observed eMSTOs to 
illustrate the capacity and limitations of isochrone clouds in explaining this 
observed phenomenon. So far, pulsational wave mixing leading to larger core 
masses has been ignored in cluster studies. Here, we remedy this lack in the 
current exploratory paper.

\section{Interior mixing and the concept of isochrone-clouds}

\subsection{Interior mixing}
The interior mixing profile results from the transport of chemical elements
inside the star. This profile cannot be determined from the basic equations of
stellar structure.  It is the net effect of several mechanisms (inter)acting
together in 3D to produce an overall mixing profile that is then simplified into
a 1D description \citep[e.g.,][]{Arnett2015}. Here, we are concerned with stars
that have a convective core and we denote the mixing profile as a function of
radial coordinate: $D_{\rm mix}(r>r_{\rm cc})$ with $r_{\rm cc}$ the boundary of
the convective core of the star according to the Ledoux criterion of convection.
As mentioned in Section \ref{section:intro}, physical processes that contribute 
to the overall mixing profile are numerous.

Near-critical rotation and its geometrical consequences in the CMD due to
the von Zeipel effect have been used to explain the eMSTO in young
clusters. Evidence for such an interpretation has been corroborated by
spectroscopically determined $v\sin i$ measurements of cluster members
\citep[e.g.,][]{Bastian2016,Dupree2017,Marino2018a,Marino2018b}. This has
motivated eMSTO fitting by using stellar evolution models based on
rotationally-induced mixing \citep[e.g.,][]{Brott2011,Lagarde2012,Milone2018}.  Early
versions of such models had been extensively studied to explain several observed
phenomena in isolated field stars \citep[e.g.,][]{Talon1997}, binary systems
\citep[e.g.,][]{deMink2009}, and stellar populations
\citep[e.g.][]{Chaboyer1995}. Such models help improve the modelling of the eMSTO
of the youngest clusters \citep[e.g.,][]{Niederhofer2015}. However, rotation
alone cannot explain the morphology observed for clusters of intermediate age
between a few hundred million years and a few giga-years
\citep[e.g.,][]{Goudfrooij2017}. Those clusters older than a giga-year 
have turnoffs in the mass range $\sim\!1.4\,$M$_\odot$ \citep{Goudfrooij2018} 
and can be well interpreted by adding the phenomenon of magnetic braking due 
to their convective envelope to rotational mixing \citet{Georgy2019}.

Stars born with masses increasing from $\sim\!$1.3 to $\sim\!$1.8\,M$_\odot$ not
only lose their convective envelope but they also transfer from a radiative core
to a well-mixed convective core, which contains between $\sim\! 10-30\%$ of the
total mass, depending on the level and efficiency of CBM.  Moreover, stars with
a well-developed convective core are subject to a whole spectrum of dissipative
IGWs, causing pulsational wave mixing throughout the radiative envelope of the
star \citep{Rogers2017}.  Space asteroseismology by MfA delivered calibrations
for the profiles of CBM and REM, denoted here as
$D_{\rm CBM}(r_{\rm env}>r>r_{\rm cc})$ with $r_{\rm env}$ the position of the
bottom of the fully radiative envelope, and $D_{\rm REM}(r_{\rm env}<r)$,
whatever their (multiple) physical origins such as rotation, waves, tides,
etc. We define $r_{\rm env}$ to be the radial coordinate where 
$D_{\rm REM} = D_{\rm CBM}$, which is always uniquely defined in our case as 
our CBM is described by a constantly decreasing function, and our REM is 
described by a constantly increasing function. This location is dependent on 
both the amount of CBM and REM, as a small amount of CBM and large amount of 
REM will result in the smallest $r_{\rm env}$, and conversely a large amount 
of CBM and a small amount of REM will result in the largest $r_{\rm env}$. 
As such, we determine the extent of the CBM region as ${\rm d}_{\rm CBM}=r_{\rm env}-
r_{cc}$.
Space asteroseismology did not reveal any correlation between the rotation
frequency and the level of CBM or REM \citep{Mombarg2019}. This lack of
correlation between rotation and mixing is in line with the early findings by
\citet{Aerts2014} based on ground-based asteroseismology and the nitrogen
surface abundance of a sample of OB-type field stars. Here, we investigate if and to
what extent the levels of CBM and REM, and their resulting enhanced core-masses,
as found from asteroseismology are capable in explaining the width and the 
morphology of the eMSTOs observed for young clusters.

\subsection{Stellar structure and evolution models}

In this work, we make use of the MESA stellar structure and evolution code
\citep{Paxton2011,Paxton2013,Paxton2015,Paxton2018,Paxton2019}.  As many such
codes, MESA solves the equations of stellar structure in a diffusive
approach. To this end, regions that are unstable against convection have a high
diffusive mixing coefficient such that the material can be considered as
instantaneously and fully mixed. The chemical mixing in radiative layers caused
by each considered mechanism is implemented with its own mixing coefficient. The
overall mixing profile in such layers is then the sum of the individual mixing
phenomena in the regions inside the star wherever that mechanism is active. We
point out that MESA's diffusive approach is different from codes that adopt an
advective numerical scheme for the treatment of rotation and its consequences
for mixing and angular momentum transport \citep[such as, e.g., the
Geneva/STAREVOL or CESTAM codes, cf.][respectively]{Ekstrom2012,Marques2013}.
Detailed asteroseismic comparisons between advective and diffusive treatments
of element transport in stellar models have yet to be carried out.

\begin{figure}
\centering
\includegraphics[width=\columnwidth]{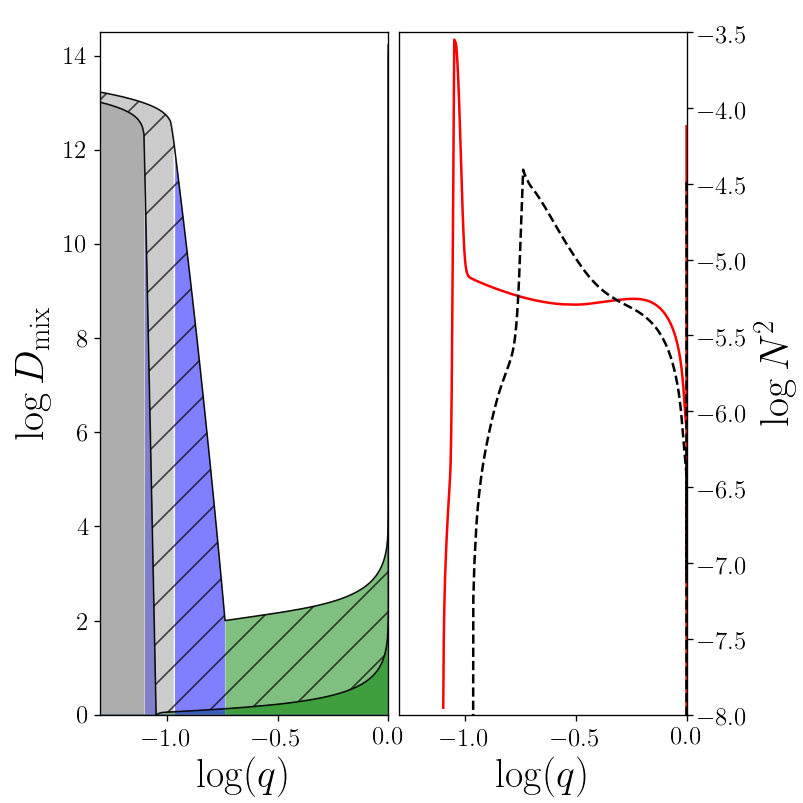}
\caption{Left: Mixing profiles for two stellar models of 3\,M$_\odot$ and
  $X_c=0.3$ for the convective core (grey), convective boundary region (CBM,
  blue) and radiative envelope region (REM, green) plotted against the logarithm
  of the mass coordinate $q=m/M_{\star}$.  The model with mixing profiles
  indicated as hatched regions has eight times higher CBM and 100 times higher
  REM than the other model, leading to an increase in core mass of 36.5\%. 
  Right: Logarithm of $N^2$ of the two models
  plotted against the logarithm of the mass coordinate $q$. The solid red line
  denotes $N^2$ of the model with a minimum amount of interior mixing, while the
  dashed black line denotes $N^2$ of the model with eight times higher CBM and
  100 times higher REM.}
\label{fig:compare_interiors}
\end{figure}

\begin{figure}
\centering
\includegraphics[width=\columnwidth]{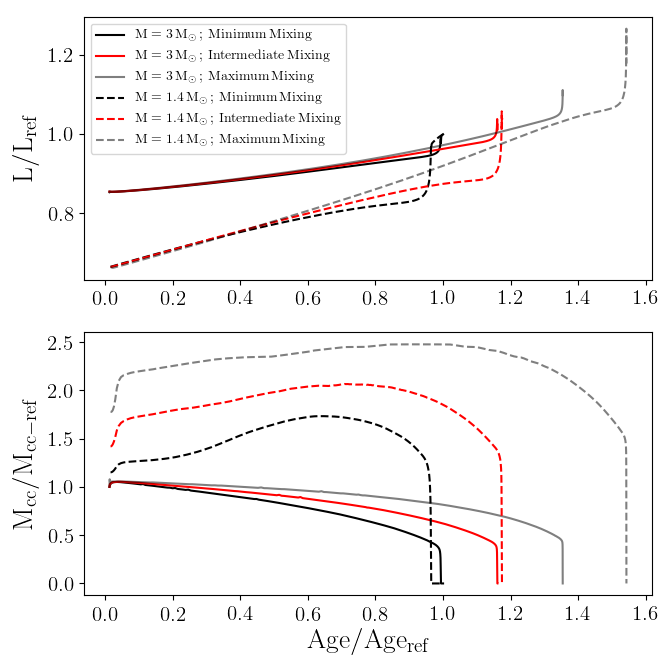}
\caption{ Top: Luminosity of evolutionary models with varied amounts of mixing
  with reference to the luminosity and age of the model with the minimum amount
  of mixing at the TAMS. Bottom: Convective core mass of evolutionary models
  with varied amounts of mixing with reference to the age of the model with the
  minimum amount of mixing at the TAMS and the core mass of the same model at
  the ZAMS.}
\label{fig:compare_mixing}
\end{figure}

We calculate {\it non-rotating spherically symmetric} equilibrium models with MESA
(version r10108) and take a pragmatic approach by considering one global free
parameter for the level of mixing for each of two areas inside the star: 1) CBM
and 2) REM.  In our scheme, CBM encompasses all forms of mixing active in the
near core regions and is approximated by the implementation of diffusive
exponential overshooting with a single free parameter, $f_{\rm ov}$, 
given by: 
\begin{equation}
    D_{\rm ov} = D_{0,cc} \exp{\frac{-2(r-r_0)}{f_{\rm ov} H_{p,cc}}},
\end{equation}
where $r_0$ is the location at which the exponential function begins, $D_{0}$ 
is the diffusive mixing coefficient at $r_0$, and $H_{p,cc}$ is the local 
pressure scale height at the boundary of the convective core \citep{Freytag1996,Paxton2011}.
The free parameter $f_{\rm ov}$ is then a scaling factor which determines how 
rapidly the exponential mixing profile decays. For the REM, we rely on a 
mixing profile due to IGW that scales with the inverse square-root of the 
density profile, as calibrated by simulations \citep{Rogers2017} and g-mode
asteroseismology \citep{Pedersen2018}.  This REM profile has an efficiency level
set at the outward end where the CBM region meets the bottom of the radiative
envelope, requiring only one free parameter.  A schematic representation of
these two mixing profiles outside the convective core can be seen in the left
panel of Fig.~\ref{fig:compare_interiors}, for a model of 3\,M$_\odot$ with
central hydrogen fraction $X_c=0.3$ for the two combinations
$(D_{\rm CBM},D_{\rm REM})=(0.005,1)$ and $(0.040,100)$ expressed in local
pressure scale height and cm$^2$\,s$^{-1}$ for $D_{\rm CBM}$ and $D_{\rm REM}$,
respectively.  This implies that we consider a global mixing profile
$D_{\rm mix}(r)$ described by two free parameters, whose observed ranges are
calibrated from asteroseismology, beyond the conventional convective core
boundary (hence for $r>r_{\rm cc}$), where $r_{\rm cc}$ is set by the Ledoux
criterion.  Thus we assume that the rotation of the cluster stars is
sufficiently slow to ignore geometrical deformation due to the centrifugal
force, which allows us to consider 1D spherically symmetric evolution
models. Following \citet{Gagnier2019}, our methodology is appropriate for stars
rotating up to about $\sim 50\%$ of their critical Keplerian rate at birth.

Aside from abundance changes, the most pronounced effect of CBM+REM is the
increase in convective core mass throughout the MS phase. In fact,
the MfA studies have shown that the g-modes are able to assess
the mass in the convective core at a certain age.  The difference in core mass
and $N(r)$ for the two example models shown schematically in the left panel of
Fig.~\ref{fig:compare_interiors} is apparent.  As can be seen, the star with
less mixing has both a less massive core (grey) and CBM region (blue) compared
to the star with more mixing (hatched). The right panel compares the Brunt-Va\"is\"al\"a
(BV) frequency (also termed buoyancy frequency) in the near-core region of these
two stars (the model with minimum interior mixing denoted by the solid red line,
the model with enhanced mixing denoted by the dashed black line). The BV
frequency, $N(r)$, is a natural frequency that occurs inside a star set by  
the local temperature gradients and 
the $\mu$-gradient, with $\mu$ the mean molecular weight.  It can be
approximated as
\begin{equation}
N^{2}=\frac{g}{H_P}\left[\delta
\left(\nabla_{\rm ad}-\nabla\right)+\varphi\nabla_\mu\right],
\label{BVformula}
\end{equation}
with 
$$
\nabla=\displaystyle{\frac{\partial \ln T}{\partial \ln P}},
\nabla_{\rm ad}=\displaystyle{\left(\frac{\partial \ln T}{\partial \ln
      P}\right)_S},
\nabla_{\mu}=\displaystyle{\frac{\partial \ln \mu}{\partial \ln P}}
$$
and 
$$
\delta=\displaystyle{\left(\frac{\partial \ln \rho}{\partial \ln
      T}\right)_{P,\mu}}, 
\varphi=\displaystyle{\left(\frac{\partial \ln \rho}{\partial \ln
      \mu}\right)_{P,T}},
$$
where $\rho$ is the density, $P$ the pressure, $T$ the temperature, and $S$ the
entropy.  Both the interior mixing properties and the shrinking convective core
cause a $\mu$-gradient in the zone left behind, affect the local behaviour of
$N(r)$. This local behaviour sets the propagation cavity of g-modes
\citep[see][for a through discussion of the probing power of such
modes]{Aerts2019}.  The difference in core mass due to CBM+REM reveals itself in
a deviating position along the evolutionary track in the CMD compared to the
position of stars with $M>1.4$\,M$_\odot$ that experience no mixing beyond the
convective core boundary, while the difference in $N(r)$ affects the properties
of the gravity modes. This the reason why asteroseismology can lead to a
high-precision estimate of the core mass for a star with detected and identified
resonant g-modes, as demonstrated by MfA.

At a given age, enhanced mixing produces a more massive core with respect to 
a star with lower levels of mixing at that age, which on the HRD in effect mimics
the evolutionary track of an initially more massive star. In other words, the
increased core mass results in an increased luminosity, radius, and temperature
with respect to the case of no mixing beyond the convective core boundary.  To
illustrate the effects of increased mixing, in Fig.~\ref{fig:compare_mixing} we
compare evolutionary tracks of a $1.4\,M_{\odot}$ and a $3\,M_{\odot}$ star,
each with three levels of mixing: 1) minimum amount of mixing, 2) an
intermediate amount of mixing (factor four enhancement in CBM and factor 10 
enhancement in REM), and 3) a maximum amount of mixing (factor eight enhancement 
in CBM and factor 100 enhancement in REM). The age is normalised with
reference to the age at the TAMS of case 1, with a
minimum amount of mixing. Likewise, the luminosity is normalised with reference
to the luminosity at the TAMS of case 1. At the TAMS, the convective core mass
is zero so we normalise the convective core mass with reference to the
convective core mass of case 1 at the zero-age main-sequence (ZAMS).  From
Fig.~\ref{fig:compare_mixing} it is seen that enhanced CBM will cause deviations
in core mass and luminosity from those cases where no CBM mixing is considered.

\subsection{Isochrone-clouds}
\label{section:isoclouds}

The concept of isochrone-clouds has been introduced by \citet{Johnston2019a} and
\citet{Johnston2019b} in order to bridge asteroseismically calibrated
$D_{mix}\left(r\right)$ profiles and isochrone fitting of pulsating and
eclipsing binaries, respectively. We briefly discuss their construction below.

Traditionally, an isochrone is constructed by interpolating in a grid of
evolutionary tracks, all computed with the same fixed input physics yet
described by a vector of free parameters $\theta$, to a desired age, $\tau$. The
vector $\theta$ minimally consists of the birth mass and the initial chemical
composition $(X,Z)$ with $X$ the initial hydrogen mass fraction and $Z$ the
metallicity. Additionally, for more dimensional model grids, $\theta$ may also
contain, e.g., the mixing length parameter of convection, $\alpha_{\rm mlt}$,
the core overshoot parameter(s), etc.  Thus, a given isochrone is described by
the vector $\phi=<\theta_i,\tau_j>$. For the construction of our
isochrone-clouds, we followed the isochrone construction as defined by
\citet{Dotter2016}, to which we refer for details.

\begin{figure*}
\centering
\includegraphics[width=1\textwidth]{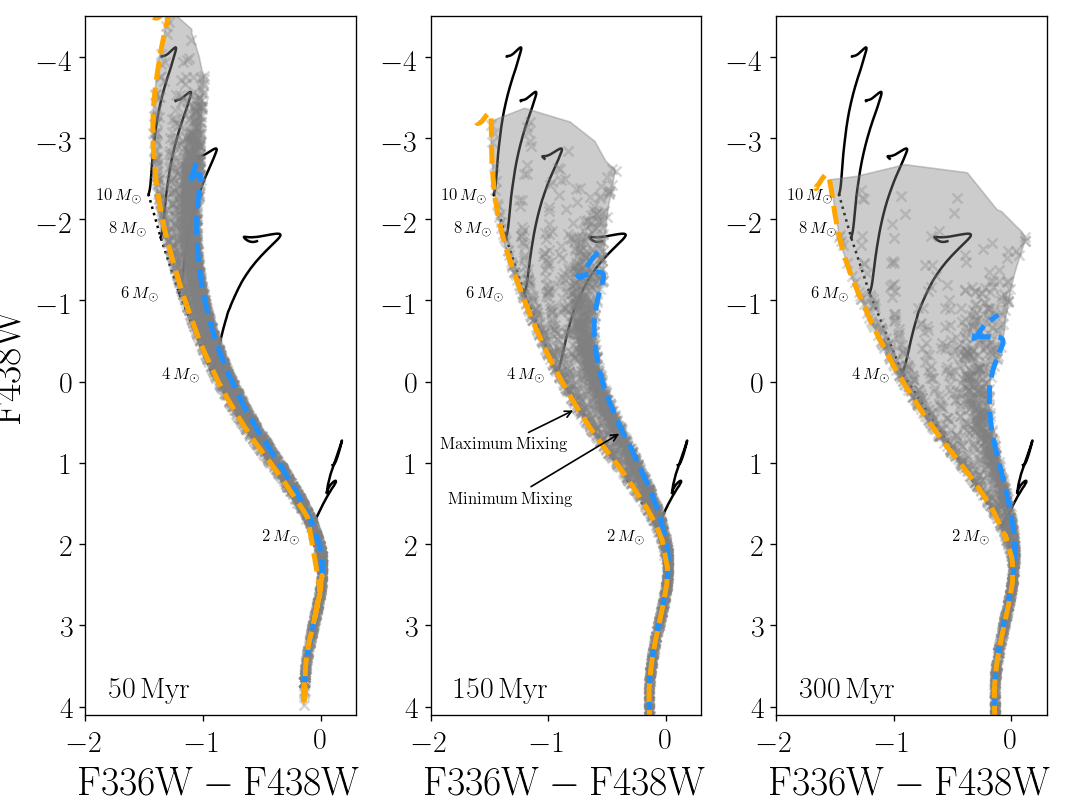}
\caption{isochrone-clouds (grey-region) at $35\,Myr$ (left), $100\,Myr$ 
  (middle), and $150\,Myr$ (right), with individual isochrones plotted 
  as grey point, and evolutionary tracks over-plotted in black. 
  The masses for the evolutionary tracks are denoted at
  the ZAMS for each track. The blue dashed line denotes the isochrone with the
  minimum amount of interior mixing and the orange dashed line represents the
  isochrone with the maximum amount of interior mixing in our model grid.}
\label{fig:compare_isoclouds}
\end{figure*}

\begin{figure}
\centering
\includegraphics[width=1\columnwidth]{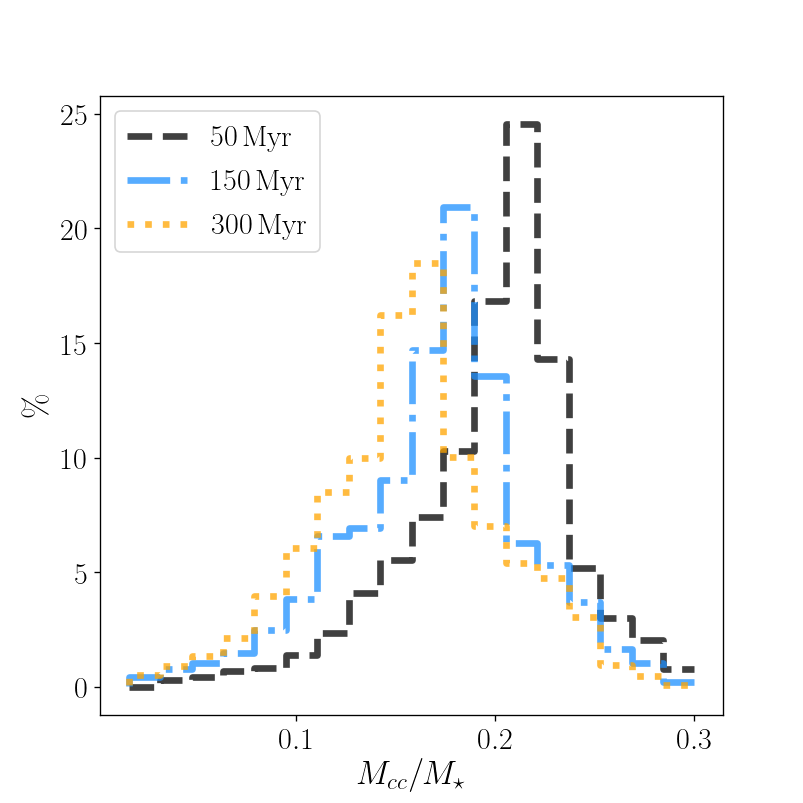}
\caption{Distribution of fractional core masses for the three isochrone-clouds
  shown in the panels of Fig.\,\ref{fig:compare_isoclouds}. The dashed black
  distribution represents the fractional core masses for the 50\,Myr isochrone-cloud,
  the blue dashed-dotted distribution is for the 150\,Myr isochrone-cloud, and
  the orange
  dotted distribution is for the 300\,Myr isochrone-cloud. }
\label{fig:icc_mass_dist}
\end{figure}

In this work, we fix $X=0.71$ and adopt the solar chemical mixture as in
\citet{Asplund2009} while relying on the OP opacity tables \citep{Seaton2005}. 
The vector $\theta_i$ corresponds to choices for the input parameters 
$(M,Z,D_{\rm CBM},D_{\rm REM})$, the latter two components being the free 
parameters that go into the computation of the overall mixing profile 
$D_{mix,i}\left(r\right)$ beyond the convective core boundary at 
$r_{\rm cc}$. 
For stars with a well-developed core, the mixing in the convective core is
instantaneous and the value of $\alpha_{\rm mlt}$ does not matter for the core
region, but it does so for the thin convective envelope for stars with masses
below some 2.5\,M$_\odot$ (higher-mass stars have pure radiative envelopes). 
The calibration of $\alpha_{\rm mlt}$ has been the subject of much research, 
with a commonly accepted calibration being $\alpha_{\rm mlt}=1.8$ \citep{Joyce2018}.
As such, we fix $\alpha_{\rm mlt}$ to this value.

An isochrone $\phi_{i,j}$ at a given age $\tau_j$ is then a single
function in the parameter space, with each effective temperature corresponding
to a single surface gravity, or luminosity. A comparison of isochrones
$\phi_{i,j}$ and $\phi_{i+n,j}$ with a factor eight increase in CBM efficiency
can be seen by comparing the orange ($\phi_{i,j}$) and blue ($\phi_{i+n,j}$)
isochrones in the panels of Fig.~\ref{fig:compare_isoclouds}.

An isochrone-cloud is then the region in the HRD or CMD at a given age,
$\tau_j$, covered by all of the isochrones $\phi_j=\sum_i \phi_{i,j}$ spanning
the range of internal mixing values, but which still have a single metallicity. 
In practice an isochrone-cloud is then just the collection of the isochrones 
$\phi_j$. Examples of isochrone-clouds are seen in grey
in the panels of Fig.~\ref{fig:compare_isoclouds}, where we note that a given
isochrone-cloud is always bound by the isochrones with the most (orange
isochrone) and least (blue isochrone) amount of interior mixing
considered. Following the panels of Fig.~\ref{fig:compare_isoclouds}, we 
see how the TAMS gradually expands in both colour and magnitude as the age
of the cloud increases, naturally producing the same morphology as is 
observed in the eMSTO. This spreading is driven by the spread in core masses
of the stars in the population with different amounts of mixing, and 
hence different levels of enhancement. An example of the evolution of
the fractional core mass distribution for each of the isochrone-clouds shown
in Fig.~\ref{fig:compare_isoclouds} is seen in comparing the distributions
in Fig.~\ref{fig:icc_mass_dist}. We notice that as the population ages, 
the fractional core mass distribution shifts to lower values and broadens.

As stated, we implement CBM as diffusive exponential overshooting in the
definition by \citet{Paxton2011}, to which we refer for details. We adopt the
range of overshooting values calibrated by asteroseismic studies as:
$f_{CBM}\in\,0.005-0.040$. We implement REM as a scaled profile described by
\cite{Pedersen2018} and based on \citet{Rogers2017} with the transitional mixing
value being: $\log\left(D_{\rm REM}\right)\in\,0-4$, corresponding with a mixing
efficiency between 1 and 10\,000\,cm$^2$\,s$^{-1}$ at the bottom of the
radiative envelope.  As said, such levels have been assessed by g-modes in
{\it Kepler\/} space photometry of core-hydrogen burning stars with a convective
core, covering the ZAMS to the TAMS.
$D_{\rm mix}(r)$ profiles calibrated by g-modes are not
yet available for hydrogen-shell burning stars. Space photometry that might be
suitable to derive them is currently being gathered by TESS \citep{Bowman2019b}.
As such, these profiles are fine to fit the eMSTO but need to get calibrated for
post core-hydrogen burning phases for future fitting of the entire CMDs of
clusters, including blue supergiants, rather than just the MS as we
do here.

Our grid of isochrone-clouds covers birth masses, 
$M_\star\in [1.2,25.0]\,$M$_\odot$, and three metallicities
$Z=0.006, 0.010, 0.014$, along with the ranges of 
$D_{\rm CBM}\in [0.005,0.040]$ and $\log D_{\rm REM}\in [0,4]$ 
derived from asteroseismic modelling of gravity-mode pulsators 
observed by the {\it Kepler\/} mission.  This grid covers 
stellar models with convective core masses in the range 
$M_{\rm cc}/M_\star\in [0,35]\%$, where a value of typically 
$\sim\!10\%$ occurs for ZAMS models that have no CBM or REM. As 
said, we limit the range in levels and shapes of mixing to those
derived from asteroseismic modelling of galactic field single 
and binary stars of intermediate and high mass, mainly but not 
exclusively from {\it Kepler\/} space photometry --- see MfA.

\section{Application to NGC\,1850 and NGC\,884}

\subsection{Fitting procedure}

Our isochrone-clouds consist of absolute magnitudes and bolometric corrections
in all HST and Johnson filters. We choose to convert these to apparent
magnitude, given a distance and extinction correction, rather than alter the
data. As such, we take distance estimates to the two clusters derived from
distance modulus and extinction estimates in the literature, yielding
$d=42\;658\,pc$ for NCG~1850 \citep{Bastian2016,Correnti2017,Yang2017,Yang2018}
and $d=2250\,pc$ for NGC~884 \citep{Kharchenko2013,Li2019,Beasor2019}. We
calculate extinction coefficients for each filter using the York Extinction
Solver \citep{YES2004}, assuming $R_v=3.1$.

Since we are concerned with fitting the eMSTO specifically for stars with a
convective core, and not the general morphology of the cluster in the whole CMD,
we make a cut in magnitude and colour for both clusters. For NGC~1850 we cut at
$m_{F439W}>19$ and $m_{F336W}-m_{F439W} < 0$, in accordance with
\citet{Bastian2016}, and for NGC~884 we cut at $G_{rp}<14$ and
$G_{bp}-G_{rp}<0.75$ following and using the Gaia DR2 data from \citet{Li2019}.

We fit isochrone-clouds to the selected part of the CMD of each cluster,
assuming that their members belong to one coeval population of single stars, at
every age between 5 and 500 Myr (in steps of 2.5 Myr between 5-50 Myr and steps
of 10 Myr between 50 and 500 Myr). Taking into account that the two free
parameters of $D_{\rm CBM}$ and $D_{\rm REM}$ are correlated with each other and
with the (core) mass and age of the star, we use the Mahalanobis Distance (MD
hereafter) as a merit function, rather than a $\chi^2$ \citep[we refer to][for a
definition and thorough discussion of its advantage in estimation of correlated
parameters]{Aerts2018}. For every cluster member, we calculate the MD with
respect to every point in the isochrone-cloud, and select the point with the
smallest MD as the best fit for that cluster member. Then, the MD for each
cluster member is summed for each isochrone-cloud, to arrive at a total fit
quality. We take the isochrone-cloud with the lowest total MD as the best fit
isochrone-cloud for the cluster.

So far, no suitable high-precision space photometry is available to perform
ensemble asteroseismic modelling of stars in young (Myrs old) open clusters.  In
absence of such data, we used the ensemble of $D_{\rm CBM}$ and $D_{\rm REM}$
estimated from asteroseismology of isolated field stars in the Milky Way in MfA as
being representative for the occurring $D_{\rm mix}(r)$ profiles inside single
stars, as discussed above.  With the aim to check how much of the width of
measured eMSTOs can be explained with a single coeval population, we applied
isochrone-cloud eMSTO fitting to two young clusters of different metallicity,
one of which hosts numerous pulsators for which ground-based multi-colour
photometry is available.

\subsection{NGC\,1850}

We use the HST, spectroscopic, and Str\"omgren data of the LMC cluster NGC\,1850
from the studies by
\citet{Bastian2016,Bastian2017,Correnti2017,Piatta2019}. These imply $Z=0.0061$,
an eMSTO-based age of $\sim 80\,$Myrs and a turn-off mass between 4.6 and
5\,M$_\odot$ \citep{Bastian2017}. This places the eMSTO stars straight into the
instability strip of the SPB stars.  No suitable time series data is available
to assess the variability of the cluster members.  The bluest MS stars of the
eMSTO of this LMC cluster have been interpreted in terms of slowly rotating
single stars \citep{DAntona2017} or low-mass-ratio binaries \citep{Yang2018}. A
major fraction of the eMSTO stars reveal H$\alpha$ emission and are therefore
interpreted as being fast rotators, but quantitative values of their $v\sin i$
are lacking so it is not possible to estimate at what fraction of the critical
rate they rotate \citep{Bastian2017}.

\begin{figure}
\centering
\includegraphics[width=9cm]{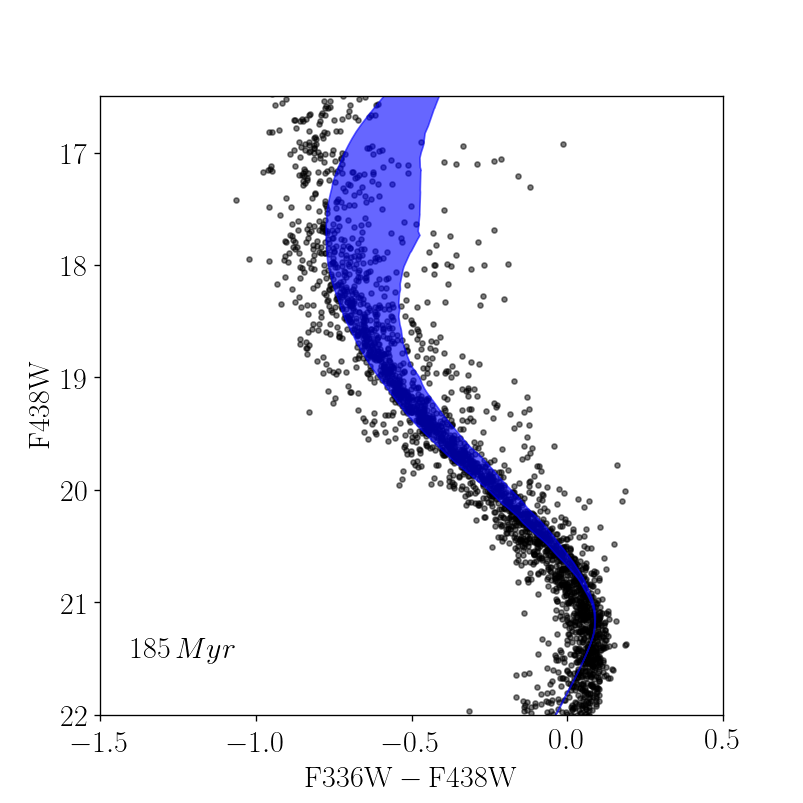}
\caption{Isochrone-cloud (blue) resulting from the best eMSTO fit for NGC\,1850 and
  representing a coeval population of single stars with an age of
  $\sim\!$185\,Myrs.}
\label{fig:isocloud_NGC1850}
\end{figure}

\begin{figure*}
\centering
\includegraphics[width=2\columnwidth]{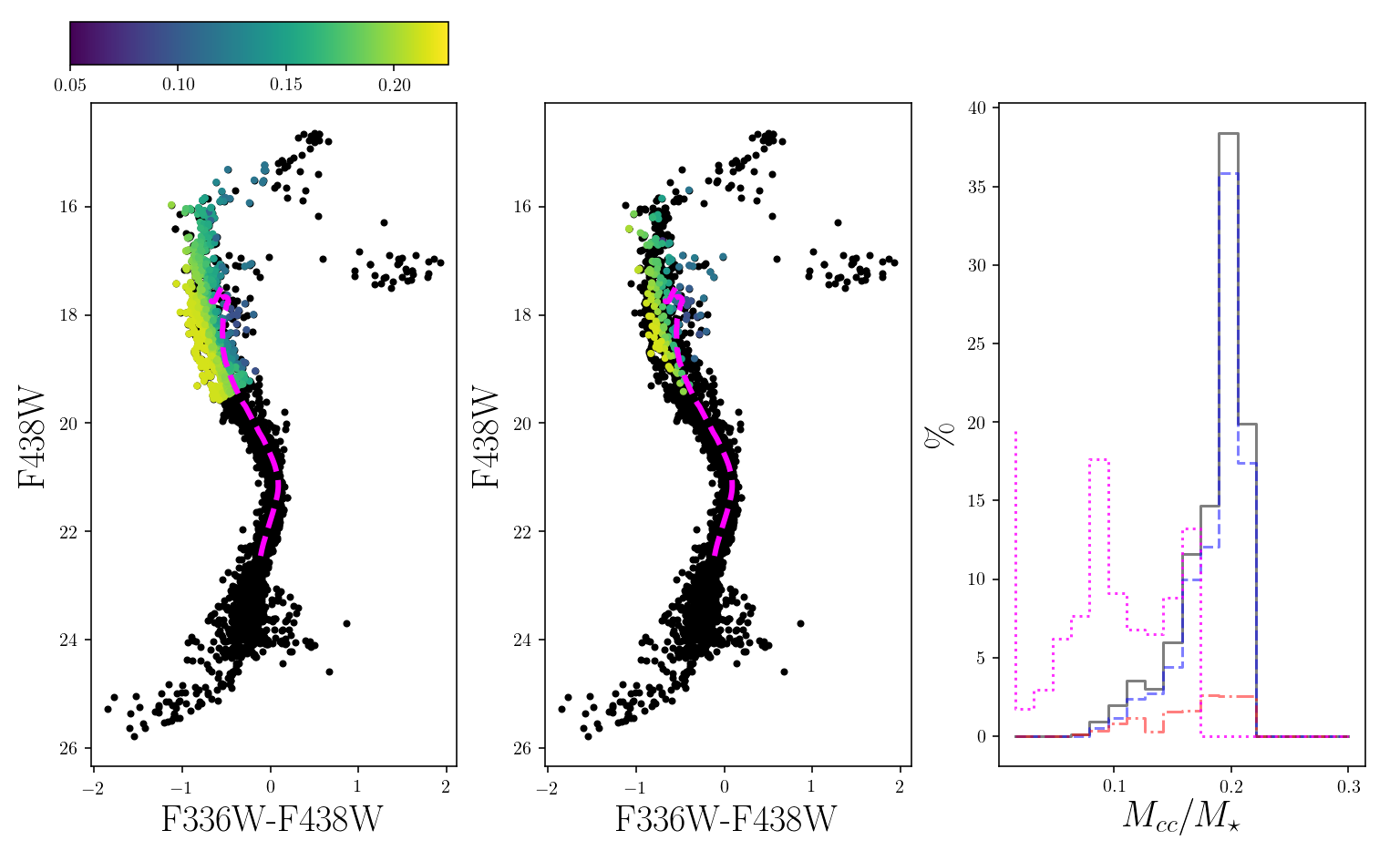}
\caption{Left: Observed stars in NGC~1850 without $H\alpha$ emission,
  colour-coded according to the ratio of the convective core mass versus total
  stellar mass.  Middle: Observed stars in NGC~1850 with $H\alpha$ emission
  colour-coded according to convective core mass. The isochrone with minimum
  amount of mixing at 185 Myr is plotted in magenta for reference in both
  panels. Right: Distribution of the fitted core mass as a fraction of total
  stellar mass, indicated in black for the full sample, in dashed blue for stars
  without $H\alpha$ emission and in dashed-dotted red for stars with $H\alpha$
  emission. The distribution of fractional core mass for the 185 Myr isochrone
  with minimum mixing is shown in dotted magenta.}
\label{fig:fit_NGC1850}
\end{figure*}

We fitted each of the stars above the colour-magnitude cut-off with our
isochrone-clouds at fixed $Z=0.006$, while letting the parameters for CBM and
REM free in the intervals $D_{\rm CBM}\in [0.005,0.040]$ and
$\log D_{\rm REM}\in [0,4]$. This results in a best fit for the age estimate of
$\sim\!$185\,Myrs, as illustrated in Fig.\,\ref{fig:isocloud_NGC1850}.  As can
be seen, the fit is quite appropriate for the eMSTO regime.  The resulting age
is $\sim 32\%$ higher than the recent age estimation based on rotating Geneva
models \citep[see][]{Bastian2017}.  The distributions of the (core) masses for
the best isochrone-cloud fit are shown in Fig.\,\ref{fig:fit_NGC1850}. The
masses cover the range $M\in [1.9,4.0]\,$M$_\odot$, which is lower than the
turn-off mass reported by \citet{Bastian2017}. This is due to the higher
fractional core masses considered in our stellar models.  The positions in the
CMD are split up in the two leftmost panels of Fig.\,\ref{fig:fit_NGC1850}, which
contain stars with H$\alpha$ in absorption (leftmost panel) and stars with
H$\alpha$ in emission (middle panel), following this spectro-photometric
information from \citet{Bastian2017}.  There is no obvious difference in
position between these two populations of stars in the sense that they occur
across the entire eMSTO, as already pointed out in \citet{Bastian2017}. We note
that while our isochrone-clouds explain a large portion of the morphology
of the eMSTO, they still struggle to explain the apparent split MS, i.e. the 
blue-most points. It has been suggested that this split-MS could be the product
of a bi-modal rotation distribution \citep{DAntona2015}, however, to date
there have not been any $v\sin i$ measurements of those stars in this region to
confirm or contrast this statement. Alternatively, we cannot rule out the possibility
that a more sophisticated fitting methodology which optimises distance and extinction 
simultaneously with age could more accurately replicate the apparent split-MS instead.

The right panel of Fig.\,\ref{fig:fit_NGC1850} shows the fractional core mass
distribution (hereafter FCMD) for the total sample (black), the sample of stars
with $H\alpha$ emission (red), without $H\alpha$ emission (blue), and the FCMD
for those points along the isochrone with the minimum mixing in our grid
(pink). The FCMD for NGC~1850 is a one-tailed skewed distribution ranging from
0.06 to 0.23 with a clear maximum at a fractional core mass of 0.2 and a sharp
drop-off at higher values.  The overall FCMD (black) is significantly shifted
with respect to the one for the case of minimal mixing (pink-dotted), which
represents models with standard core masses without CBM along the evolution.  It
can also be seen from the three panels that the distribution of
$M_{\rm cc}/M_\star$ is uncorrelated with the H$\alpha$ behaviour, while being
cleanly segregated in the sense that stars with higher fraction of
$M_{\rm cc}/M_\star\%$ (i.e., stars with a high amount of interior mixing) are
situated on the blue side of the eMSTO.  Our findings imply that the position in
the eMSTO is connected with the relative ratio of the core versus overall
stellar mass, and not with the surface rotation of the stars. This is not
necessarily in contradiction with previous interpretations of stellar rotation
being the dominant cause of the eMSTO, as our way of increased near-core and
envelope mixing is not coupled to a specific physical process. Our findings show
that efficient interior mixing, as it has been calibrated from asteroseismology
for field stars, is sufficient to explain a large portion of the observed width
of the eMSTO for non-rotating 1D evolutionary models. Our interpretation does
not require stars to be rotating fast, and particularly not near critical, but
rather suggests that an efficient mixing mechanism is at work in the near-core
region of the star. The cause of this CBM could be rotation, or other
phenomena such as waves, tides, etc.

\subsection{NGC\,884}
\label{section:results_ngc884}

NGC\,884 is a bright northern galactic open clusters that has been observed in
ground-based months-long time-series multi-colour photometry with the aim of
detecting stellar oscillations in the cluster members \citep{Saesen2010}.  This
extensive monitoring was done in an attempt to perform cluster ensemble
asteroseismology \citep{Saesen2013}. The campaign revealed 115 detected
oscillation frequencies in 65 B-type stars in the cluster, covering amplitudes
in the V band between roughly 0.5 and 18\,mmag \citep[Fig.6
in][]{Saesen2013}. While those data are far less precise than {\it Kepler\/} and
TESS space photometric time series and not suitable to perform asteroseismic
modelling to derive $D_{\rm mix}(r)$ for individual stars, this cluster ensemble
of B-type pulsators did lead to an asteroseismic age range of $\sim\! 13.2-19.1$
Myrs, based on the detected pressure modes in eight $\beta\,$Cep stars in the
cluster. This result was obtained from stellar models with penetrative
convection computed with the code CL\'ES \citep{Scuflaire2008}.

The presence of numerous non-radial pulsators with a $v\sin i$ measurement in
NGC\,884 makes it one of the optimal clusters that lends itself to test if its
eMSTO bears any correlation with the oscillation properties of its members,
rather than or in addition to their surface rotation.  Moreover, in such a young
cluster, only very efficient mixing processes that act on a timescale much
shorter than the nuclear time scale will be able to build up a larger core mass
due to CBM since the ZAMS. We test these hypotheses from isochrone-cloud fitting
allowing for $\log D_{\rm REM}\in [0,4]$ as calibrated from MfA.

\begin{figure}
\centering
\includegraphics[width=9cm]{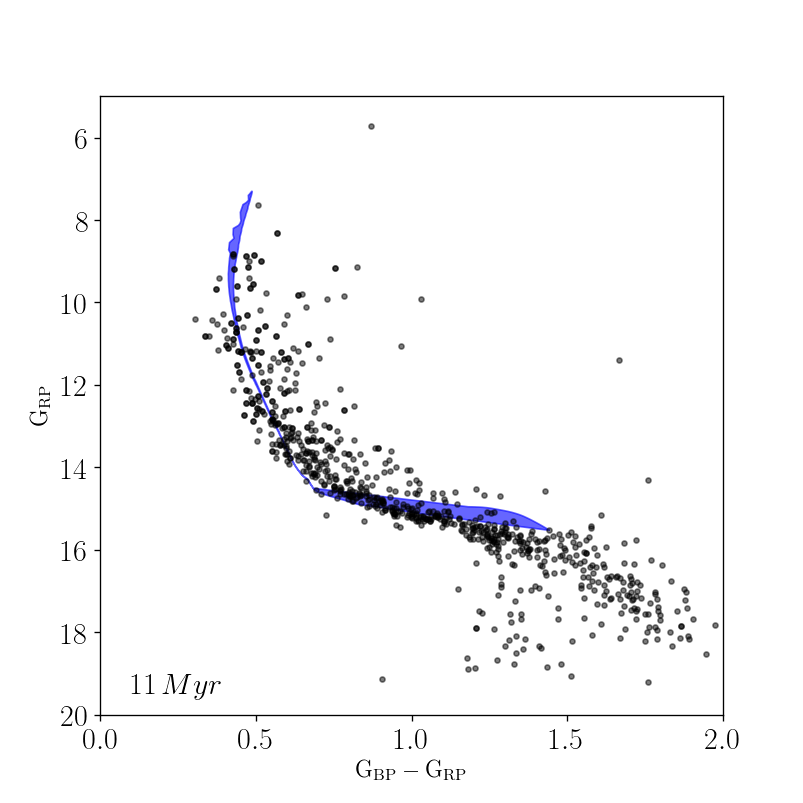}
\caption{Isochrone-cloud (blue) resulting from the best eMSTO fit for NGC\,884,
  representing a coeval population of single stars with an age of
  $\sim\!$11\,Myrs.}
\label{fig:isocloud_NGC884}
\end{figure}

We fitted the observed eMSTO in the Gaia DR2 CMD used by \citet{Li2019} with our
isochrone-clouds, again assuming one coeval population of single stars. This
time we fixed $Z=0.014$, given the occurrence of numerous multiperiodic B-type
pulsators in the cluster. The excitation of stellar oscillations in such stars
is based on the $\kappa$ mechanism. For this mechanism to work and excite
pressure (p-) modes in $\beta\,$Cep stars, a $Z-$ (opacity) bump due to the
iron-group elements must occur in the outer envelope of such stars, demanding
$Z>0.010$ for the adopted solar mixture \citep{Asplund2009} and OP opacity
tables \citep{Miglio2007}. 

Our isochrone-cloud fit results in an age estimate of $\sim\!11$\,Myrs and a
stellar mass coverage of $M\in [1.7,17.8]\,$M$_\odot$, in agreement with
\citet{Saesen2013} and with the fact that $p-$ and $g-$mode oscillations were
detected in cluster members \citep{Saesen2013}.  This age estimate is a
slightly younger than the 14\,Myr estimate by \citet{Li2019}, who used the
SYCLIST suite of Geneva rotating stellar models \citep{Georgy2014}. However, 
given the different model assumptions and fit uncertainties, we consider these
to results to be in agreement. Considering such model differences, our estimate
is also in agreement with those of 13.2-19.1 Myr produced by \citet{Saesen2013}, 
who ignored the effect of rotation on oscillation frequencies and made assumptions
on mode identification. The result
of our fit are shown in Fig.\,\ref{fig:isocloud_NGC884} and is to be compared
with Fig.\,7 in \citet{Li2019}, who showed the case of a non-rotating population
as well as one rotating at 90\% critical. 
% \citet{Li2019} suggest that an age of
% 30~Myr or more would be unlikely as this would imply the existence of a
% population of White Dwarfs, which has yet to be observed. However, the presence
% of such a population depends on stars in the correct mass range to have evolved
% to produce such evolutionary products. Our modelling indicates that the most
% massive stars are still above $9\,M_{\odot}$, implying that any evolutionary
% end-products would be undetectable in the CMD.

\begin{figure*}
\centering
\includegraphics[width=2\columnwidth]{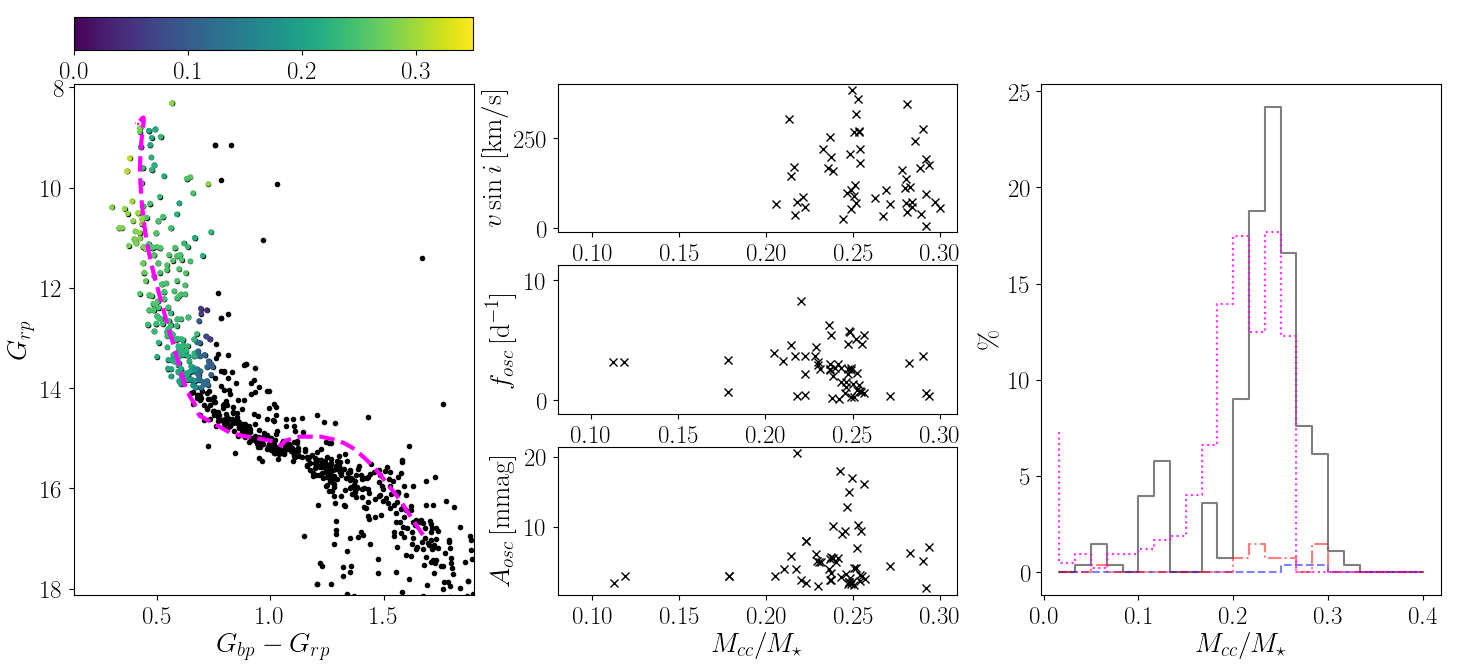}
\caption{Fit results for isochrone-cloud with $Z=0.014$, leading to age of
  $\sim\!11$\,Myrs.  Left: Same as left panel of Fig.~\ref{fig:fit_NGC1850}. The upper,
  middle and lower panel of the central plots show $v\sin i$, the dominant
  oscillation frequency and its amplitude vs.\ fractional core mass. Right: Same as right
  panel of Fig.~\ref{fig:fit_NGC1850}, but blue denotes stars with measured
  oscillations and red denotes stars with measured $v\sin i$. Pink dotted distribution
  is for those stars with the minimum amount of mixing.
  }
\label{fig:results_NGC884}
\end{figure*}

The results for the estimates of the core masses are shown in the left panel of
Fig.\,\ref{fig:results_NGC884}. The central panels of the figure show the stars in
the cluster with measurements of $v\sin i$ (upper), and the frequency and
amplitude of the dominant oscillation mode (middle and lower) as a function of
the fractional core mass. We check all of these for statistically significant
correlations via linear regression modelling, with the results listed in 
Table~\ref{tab:regression_results}. We find no significant correlations 
between either $v\sin i$, $f_{osc}$, or $A_{osc}$ and the fractional core masses.

The total FCMD (solid black distribution in
the right panel of Fig.~\ref{fig:results_NGC884}) ranges between 0.04 and 0.32,
with a clear maximum at a fractional core mass of 0.25 and an extended
population of stars with higher fractional core masses. The fractional core mass
distributions for those stars with $v\sin i$ estimates (red dashed-dotted
distribution) and with $f_{osc}$ measurements (blue dashed distribution) are
similar and show no clear trends. The resulting fit distribution is different
from the distribution of fractional core masses assuming a population
with the least amount of interior mixing in the grid (pink-dotted distribution),
indicating that also in this much younger cluster members, a significant amount
of mixing occurs for its eMSTO stars. Pulsational wave mixing by IGW qualifies
as an explanation, just as it provides a suitable mechanism for efficient AM
transport \citep[cf.,][and references therein]{Aerts2019,Bowman2019b}. Given
that the stars at the eMSTO of NGC\,884 are more massive than those for
NGC\,1850, and that the amplitude of detected IGW was found to be proportional
with the stellar mass and luminosity and independent of metallicity
\citep{Bowman2019b}, IGWs are predicted to be more efficient for more massive
stars.  This does not exclude other phenomena that could also act on the
required short time scale to enlarge the core masses of the eMSTO stars. 

\begin{table}
  \caption{$R^2$- and $p$-values 
of statistical regressions for various stellar parameters with the derived 
    fractional core mass $M_{cc}/M_{\star}$.}
\label{tab:regression_results}
  \centering
  \begin{tabular}{lcc}
  \hline\hline
  y-variable &  $R^2$   &  $p$  \\
  \hline
  $v\sin i$  & 0.007 & 0.55 \\
  $f_{osc}$  & 0.003 & 0.67 \\
  $A_{osc}$  & 0.034 & 0.16 \\
  \hline
  \end{tabular}
  \end{table}

\section{Discussion and Conclusions}
\label{section:discussion}

NGC~1850 and NGC~884 are two young open clusters known to have significantly
different age and metallicity. Their eMSTO is situated in the range of
intermediate-mass B-type stars.  From isochrone-cloud fitting, based on interior
mixing levels as derived from asteroseismology of single field stars of similar
mass, we find that the stars at the eMSTO of both clusters have enhanced core
masses compared to standard models without core boundary and envelope mixing. As
a result of the interior mixing, two stars with the same age and similar birth
mass, but different interior mixing levels would appear in different locations
on the CMD, where the star with enhanced mixing would be located blueward of its
less mixed analogue.  

From our isochrone-cloud fitting, we derive an age estimate of $11\, Myr$ for
NGC~884 and $185\,Myr$ for NGC~1850. We find that both of these estimates agree
with previous estimates considering the differences in model physics and fitting 
methodologies. The relative core mass fractions 
reveal a broader distribution for the younger and more metal-rich cluster NGC~884
than for the older LMC cluster NGC~1850. This suggests that (a) CBM mechanism(s)
are active on a time scale much shorter than the nuclear time scale and that the
mechanism(s) are more effective as the stellar masses are higher. 

Our isochrone-cloud modelling is based on spherically-symmetric stellar models
with enhanced core masses compared to standard 1D models without extra
mixing. This enables us to explain the morphology of the eMSTO of the two
clusters to a large extent. Just as with asteroseismic modelling of single and
binary pulsators, the cause of the interior mixing was left open and could be
due to rotation, pulsations, waves, tides, binary merging, etc. The larger the
CBM, the higher the relative core mass fraction gets during the first part of
the MS, so the bluer the stars occurs in the HRD compared to models without
interior mixing. We do not find any relationship between the $v\sin\,i$ of the
stars and the core mass fraction.

With our work, we provide an entirely different explanation for the eMSTO than
studies in the recent literature, which rely on models of stars with a spread of
rotation rates, at least some of which are rotating near their critical velocity 
(as corroborated by the observation of $H\alpha$ emission, with the assumption that
such emission is caused by near critical rotation) such that colour ($T_{\rm eff}$)
modifications due to geometric distortion and the Von Zeipel effect come into play
\citep[e.g.,][]{Georgy2019,Li2019,Gossage2019}. Rapid rotation and a high level
of interior mixing are not mutually exclusive and both effects may be active in
practice. However, it seems unrealistic to assume that almost all stars at the
eMSTO rotate near critical. To that end, a sub-population of stars rotating at near
critical rotation is still a candidate to explain the split-MS phenomenon. 
For NGC~884, high-precision spectroscopy revealed
that the fastest rotating emission-line stars at the eMSTO rotate at about half
their critical rate rather than near critical. For this cluster, the Von Zeipel
effect hence does not offer a good explanation.

Our results points towards the need of higher core masses in stellar models for
stars born with a mass above $\sim\!1.4\,M_{\odot}$.  These higher core masses
can be caused by enhanced amounts of interior mixing, which results in higher
luminosities compared to the case where no extra mixing is considered in the
near-core region.  Mixing induced by the overall effect of penetrative
convection, core overshooting, resonant standing gravity modes and/or
dissipative IGW, is naturally expected for stars with masses higher than
$1.4\,M_{\odot}$, irrespective of their rotation rate.  This ``cutoff'' mass has
been found observationally from eMSTOs of numerous clusters
\citep{Goudfrooij2018}. Rotational mixing due to meridional circulation and
other instabilities induced by rotation \citep{Heger2000,Aerts2019} would not
necessarily lead to such a strict cutoff in mass. In order to explain this
observed cutoff from rotating models, one can rely on magnetic braking induced
by the outer convective envelope of stars \citep{Georgy2019}. Our mechanism, on
the other hand, focuses on the innermost regions of the stars rather than on
their envelope behaviour to explain the eMSTO and the cutoff at $1.4\,M_{\odot}$
for the different eMSTO morphology in observed clusters.

We conclude for both considered clusters that isochrone-clouds can reproduce to
a large extent the morphology of their eMSTO in terms of interior mixing.
Although we fit only the eMSTO, the dimmer and redder regions of the CMD are
adequately described by the best isochrone-cloud model, despite having had no
influence on the fitting procedure. This suggests that our models are well
calibrated. However, our tentative conclusions must be tested on a large sample
of young open clusters covering a good spread in age, metallicity, and turn-off
mass. Ideally, this work will be extended to incorporate space-based asteroseismology
of YMCs to be delivered by TESS. We plan to take up such more systematic studies 
in future work.

\section*{Acknowledgements}
We thank the referee for their comments which improved the manuscript.
This work sparked from discussions among the authors during the 2018 Lorentz
Workshop ``Weighing Stars from Birth to Death: How to Determine Stellar
Masses?''; we are grateful to the staff of the Lorentz Center for the kind
hospitality and support. We are grateful to Chengyuan Li for having provided us 
with the Gaia CMD data from his paper on NGC\,884 in electronic format.
%and to Aaron Dotter for valuable suggestions prior to submission.  
The research leading to these results has received funding from the European 
Research Council (ERC) under the European Union’s Horizon 2020 research and 
innovation programme (grant agreements N$^\circ$670519: MAMSIE with PI Aerts and N$^\circ$646928: Multi-Pop with PI Bastian) and from the Research Foundation 
Flanders (FWO) under grant agreement G0A2917N: BlackGEM with PI Aerts. The 
computational resources and services used in this work were provided by the 
VSC (Flemish Supercomputer Center), funded by the Research Foundation - 
Flanders (FWO) and the Flemish Government – department EWI to PI Johnston.

\bibliographystyle{aa}
\bibliography{accepted_v1.bib}

%\bsp
\label{lastpage}

\end{document}